\documentclass[sigconf]{acmart}
\usepackage[utf8]{inputenc}
\usepackage{newunicodechar}
\newunicodechar{～}{\textasciitilde}
\renewcommand\footnotetextcopyrightpermission[1]{}

\makeatletter
\renewcommand\@formatdoi[1]{\ignorespaces}
\AtBeginDocument{%
  }

\setcopyright{acmlicensed}
\copyrightyear{2025}
\acmYear{2018}
\acmDOI{XXXXXXX.XXXXXXX}
\acmConference[Conference acronym 'XX]{Make sure to enter the correct
  conference title from your rights confirmation email}{June 03--05,
  2018}{Woodstock, NY}
\acmISBN{978-1-4503-XXXX-X/2018/06}

\usepackage{subcaption} 
\usepackage[most]{tcolorbox} 
\usepackage{graphicx} 
\usepackage{algorithm}
\usepackage{algpseudocode}
\usepackage{multirow}
\usepackage[table]{xcolor}
\usepackage{enumitem}

\definecolor{wheat}{rgb}{0.96,0.87,0.70}
\definecolor{color_blue}{rgb}{252,182,165}
\definecolor{color_pink}{rgb}{255,217,178}
\definecolor{myblue}{RGB}{0, 102, 204} 
\definecolor{myred}{RGB}{220, 53, 69} 
\definecolor{examplecolor}{HTML}{C0392B}   
\definecolor{refinedcolor}{HTML}{27AE60}

\begin{document}


\title{SpeechAgent: An End-to-End Mobile Infrastructure for Speech Impairment Assistance}

\author{Haowei Lou}
\email{haowei.lou@unsw.edu.au}
\orcid{0009-0009-1359-872X}
\affiliation{%
  \institution{University of New South Wales}
  \city{Sydney}
  \state{NSW}
  \country{Australia}
}

\author{Chengkai Huang}
\email{chengkai.huang@mq.com}
\orcid{0000-0002-1630-424X}
\affiliation{%
  \institution{Macquarie University}
  \city{Sydney}
  \state{NSW}
  \country{Australia}
}

\author{Hye-Young Paik}
\email{h.paik@unsw.edu.au}
\orcid{0000-0003-4425-7388}
\affiliation{%
  \institution{University of New South Wales}
  \city{Sydney}
  \state{NSW}
  \country{Australia}
}

\author{Yongquan Hu}
\email{yongquan@ahlab.org}
\orcid{0000-0003-4425-7388}
\affiliation{%
  \institution{National University of Singapore}
  \country{Singapore}
}

\author{Aaron Quigley}
\email{aaron.quigley@csiro.au}
\orcid{0000-0002-5274-6889}
\affiliation{%
  \institution{CSIRO’s Data61}
  \city{Sydney}
  \state{NSW}
  \country{Australia}
}

\author{Wen Hu}
\email{wen.hu@unsw.edu.au}
\affiliation{%
  \institution{University of New South Wales}
  \city{Sydney}
  \state{NSW}
  \country{Australia}
}

\author{Lina Yao}
\email{lina.yao@data61.com.au}
\orcid{0000-0002-4149-839X}
\affiliation{%
  \institution{CSIRO's Data61}
  \institution{University of New South Wales}
  \city{Sydney}
  \state{NSW}
  \country{Australia}
}

   




\begin{abstract}

Speech is essential for human communication, yet millions of people face impairments such as dysarthria, stuttering, and aphasia—conditions that often lead to social isolation and reduced participation.
Despite recent progress in automatic speech recognition (ASR) and text-to-speech (TTS) technologies, accessible web and mobile infrastructures for users with impaired speech remain limited, hindering the practical adoption of these advances in daily communication. 
To bridge this gap, we present \textbf{SpeechAgent},
a mobile \textbf{SpeechAgent} designed to facilitate people with speech impairments in everyday communication. 
The system integrates large language model (LLM)–driven reasoning with advanced speech processing modules, providing adaptive support tailored to diverse impairment types.
To ensure real-world practicality, we develop a structured deployment pipeline that enables real-time speech processing on mobile and edge devices, achieving imperceptible latency while maintaining high accuracy and speech quality.
Evaluation on real-world impaired speech datasets and edge-device latency profiling confirms that SpeechAgent delivers both effective and user-friendly performance, demonstrating its feasibility for personalized, day-to-day assistive communication.
Our code, dataset, and speech samples are publicly available\footnote{https://anonymous.4open.science/r/SpeechAgentDemo-48EE/}.
\end{abstract}
\maketitle

\section{Introduction}
\begin{figure}
    \centering
    \includegraphics[width=0.75\linewidth]{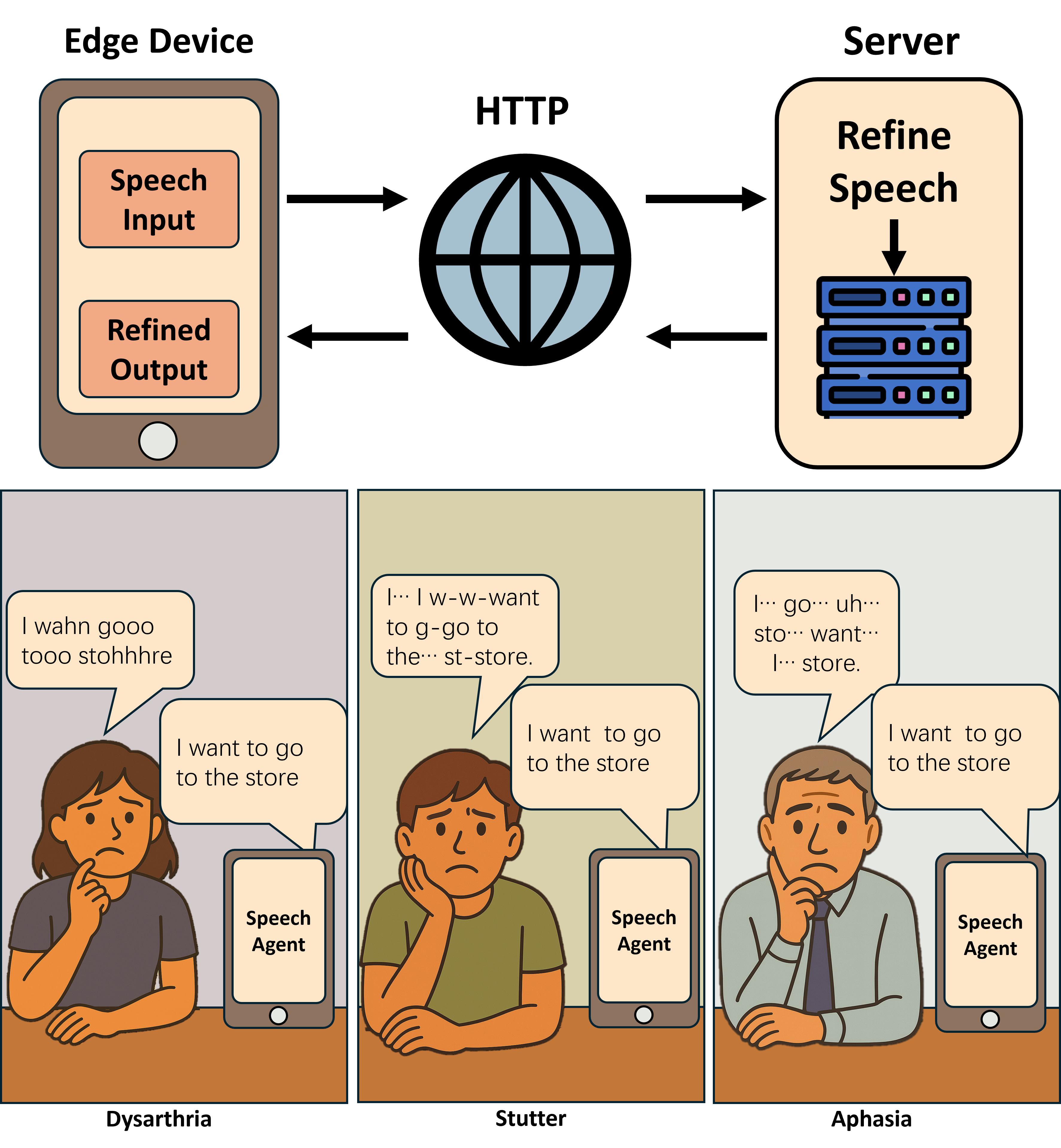}
    \caption{Overview of the \textbf{SpeechAgent} system. The edge device records impaired speech and sends it to the cloud server via HTTP, where the Refine Speech module performs recognition and restoration before returning the refined output. The examples illustrate how \textbf{SpeechAgent} converts dysarthric, stuttered, and aphasic speech into clear, fluent sentences.}    
    \label{fig:placeholder}
\end{figure}

Speech serves as the primary channel through which humans externalize thought and connect with others.
It is not only a vehicle for exchanging information but also a medium for expressing identity, intention, and emotion \cite{schuller2013computational}. 
Through speech, individuals shape relationships, share knowledge, and coordinate collective action within society. The ability to articulate ideas clearly fosters trust and collaboration, allowing people to work together toward common goals \cite{Kohler_2017,wolvin1981speech,gardiner1932theory}.
Viewed from this perspective, speech is more than the transmission of words—it is an ongoing process of aligning one’s internal thoughts with how they are perceived and understood by others, forming the foundation of self-expression and social participation.

However, not everyone can express themselves fluently through speech. Many individuals experience impairments that make their speech fragmented, unclear, or difficult to follow. Such challenges often arise from clinical speech and language disorders that directly affect articulation, fluency, or coherence \cite{hamaguchi2010childhood,eichorn2024assessment}. 
Among the most prevalent are \textit{(i) dysarthria}, which arises from motor impairments and results in slurred or slow speech due to reduced control of the articulatory muscles~\cite{enderby2013disorders}; \textit{(ii) stuttering}, characterized by involuntary repetitions or prolongations that disrupt the natural flow of speech~\cite{prasse2008stuttering}; and \textit{(iii) aphasia}, typically caused by neurological damage such as stroke, which impairs a person's ability to produce or comprehend language~\cite{damasio1992aphasia,brady2016speech}.

These disorders extend beyond the clinical symptoms, shaping how people interact and are perceived in everyday communication.
Individuals with speech impairments often struggle to make themselves understood, facing frequent communication breakdowns that result in misunderstandings, interruptions, or dismissals. Such experiences can be discouraging and might,
over time, restrict participation in social, educational, and professional contexts,
leading to isolation and reduced quality of life~\cite{mccormack2010my,bashir1992children}. Addressing such barriers is therefore essential. This motivates the design of our \textbf{mobile SpeechAgent}, designed to assist individuals with dysarthria, stuttering, and aphasia in achieving clearer and more effective communication.  




Previous research in speech and language technologies has aimed to support individuals with speech impairment. Traditional approaches include augmentative and alternative communication (AAC) devices~\cite{beukelman1998augmentative,schlosser2008effects}, rule-based grammar correction systems~\cite{sidorov2013rule,naber2003rule}, and computer-assisted therapy programs~\cite{grossinho2014interactive,chapelle2001computer}. 
While these tools have provided valuable support in structured clinical or educational domains, they are often rigid and rely heavily on predefined templates or rule mappings. Such designs perform well for predictable, domain-limited input, but they struggle with the fragmented, ambiguous, and dynamic pattern of real-world speech.

In clinical environments, technological interventions have typically emphasized long-term rehabilitation rather than live conversational repair. For instance, computer-mediated aphasia programs focus on vocabulary relearning~\cite{brady2016speech}, while stuttering management tools provide pacing strategies or fluency feedback~\cite{perez2016stuttering}. Similarly, interactive therapy software often motivates patients with corrective prompts or gamified exercises~\cite{grossinho2014interactive,rodriguez2008comunica}. Although these methods are effective in structured therapeutic environments, they are not designed to adapt flexibly to open-ended, everyday usage as well.  
Recent research has explored 
training automatic speech recognition (ASR) models to better transcribe impaired speech~\cite{vinotha2024enhancing,lea2023user, huang2025leveraging}, or developing detection system and classifying different types of speech impairments~\cite{sekhar2022dysarthric,al2022stuttering}. While these approaches improve the ability to recognize or analyze disordered speech, they are primarily diagnostic in nature. They do not provide direct communicative assistance to the user in real time, nor do they help transform impaired speech into a clearer, more comprehensible form during everyday interaction. 
To the best of our knowledge, there is currently no assistive communication tool that offers real-time refinement of impaired speech for practical, everyday communication. 
To address this gap, we introduce a mobile assistive \textbf{SpeechAgent} that functions as an assistive tool for spontaneous communication. The system refines impaired speech into a clearer form that preserves the speaker’s intention, allowing users to communicate easily in everyday conversations.

Our assistive \textbf{SpeechAgent} advances beyond traditional speech-impairment refinement pipelines\cite{cleanvoice2025stutter,arjun2020automatic} by integrating perception, reasoning, and generation into a unified agent framework. 
It combines open-sourced and custom-trained models to not only recognize disordered speech but also understand and restore the speaker’s communicative intention. It contains a robust, general-purpose ASR model~\cite{radford2023robust} transcribes speech reliably across diverse acoustic conditions, while a dedicated impairment recognition model captures speaker-specific disorder patterns, enabling adaptive, context-aware processing for downstream refinement.  Building upon these perceptual foundations, the agent employs large language models (LLMs)~\cite{achiam2023gpt,team2024gemini,yang2025qwen3,team2025gemma} as cognitive refinement engines that reason about the meaning behind fragmented, ambiguous, or disfluent utterances. 
Rather than performing surface-level grammatical correction~\cite{naber2003rule,gramformer}, the LLM reconstructs semantically coherent and faithful text that preserves the speaker’s intent. 
Finally, a natural and expressive text-to-speech (TTS) module~\cite{lou2025generalized,ren2020fastspeech} vocalizes the refined text, completing a closed perception–reasoning generation loop that transforms impaired speech into clear, intelligible, and easy to understand speech.
The main contributions of this paper are as follows:


\begin{itemize}[leftmargin=1em, itemindent=0em, itemsep=0.25em, topsep=0.25em]
    \item We propose a mobile \textbf{SpeechAgent} that leverages LLM-based reasoning and advanced speech techniques to support users with speech impairments in everyday communication.
    \item  We extend the system with mechanisms to recognize different types of speech impairments and adapt the agent's refinement strategies accordingly.
    \item We release a benchmark suite and evaluation pipeline for impaired-speech systems, combining both system-level (latency, throughput, scalability) and user-level (clarity, usability) metrics.
    \item We conduct comprehensive experiments, and our results demonstrate that the proposed \textbf{SpeechAgent} is easy to use and supports real-time response on edge devices.
\end{itemize}

\section{Related Work}
\noindent \textbf{Speech Impairments.} 
Speech impairments arise from diverse neurological, developmental, or physiological conditions and encompass a wide range of disorders affecting clarity, fluency, and intelligibility. This work focuses on three representative types: dysarthria, stuttering, and aphasia. Dysarthria is an acoustic impairment in which speech sounds are slurred, strained, or unusually slow due to weakened motor control of articulatory muscles, often resulting from stroke, cerebral palsy, traumatic brain injury, or degenerative diseases such as Parkinson’s~\cite{enderby2013disorders,pinto2004treatments}.
Stuttering, by contrast, disrupts speech rhythm through involuntary repetitions (e.g., “b-b-b-book”), prolongations (“ssssun”), or silent blocks, producing hesitation and tension perceptible even in short utterances. It has both developmental and neurological origins and is associated with atypical timing and coordination in neural circuits governing speech planning and execution~\cite{conture1990stuttering,bloodstein2021handbook}.
Aphasia differs from the above as it primarily involves linguistic and cognitive deficits in language production and comprehension~\cite{ellis1983wernicke,wiener2004inhibition,damasio1992aphasia}. Although articulation and prosody may remain intact, the semantic coherence of speech is disrupted, making it difficult for listeners to grasp the intended meaning.

\noindent \textbf{Assistive Tools for Speech Impairments.} 
Research in speech and language technologies has long aimed to assist individuals with communication difficulties. Traditional tools largely rely on rule-based correction, including speech therapy software~\cite{shipley2023assessment}, augmentative and alternative communication (AAC) devices~\cite{beukelman1998augmentative,schlosser2008effects}, and computer-assisted language learning (CALL) platforms~\cite{chapelle2001computer}. While useful within their respective domains, these systems operate on fixed rules or phrase banks, offering predictable corrections but limited adaptability. Because they depend on explicit error–correction mappings, such approaches struggle with the variability and ambiguity of spontaneous or impaired speech. In clinical contexts, technological supports have primarily targeted rehabilitation rather than real-time conversational repair. Programs for aphasia focus on relearning vocabulary and syntax~\cite{brady2016speech}, while stuttering management systems provide pacing or fluency feedback~\cite{perez2016stuttering}. Other interactive therapy tools use monitoring, feedback, or gamified exercises to motivate patients~\cite{grossinho2014interactive,rodriguez2008comunica}. Although effective in structured therapeutic environments, these systems are not designed to adapt flexibly to the open-ended nature of everyday communication.

Most prior systems either focus on therapeutic training in clinical settings or rely on rigid rule-based correction, which limits adaptability to spontaneous and dynamic communication. As a result, individuals suffering from speech impairment often lack access to tools that can flexibly support them in ordinary conversations, where clarity and intent alignment are most essential. 

\section{Problem Formulation}
Human communication can be viewed as the process of expressing an underlying communicative intention $I$ through a speech signal $S$. Formally, the speaker produces speech according to a generative process $p(S \mid I)$, while the listener attempts to recover the intention by estimating $p(I \mid S)$.

For healthy speakers, the mapping between $I$ and $S$ is generally straightforward. The speech carries sufficient information and is clear enough for listeners to infer the intention with high accuracy, $p(I \mid S)$ is well aligned with the true $I$. In contrast, for speakers with speech impairments, who produce speech with \emph{acoustic distortions} (e.g., disfluency, articulation errors) or \emph{logical distortions} degrade this mapping, making $p(I \mid S)$ less reliable and making it difficult for listeners to easily understand $I$ from $S$.

The problem of speech refinement can be formally defined as finding a transformation $f$ that maps an impaired speech $S$ to a refined version $S'$, guided by the speaker's latent communicative intention $I$. Since the true intention $I$ is unobservable, we introduce a latent variable $Z$ to represent the model’s internal understanding of the speaker’s intent. $Z$ serves as an implicit cognitive state formed by the LLM’s reasoning process, capturing semantic and pragmatic cues that guide refinement.

\begin{equation}
       S^{\ast} = \underset{S'}{\mathrm{argmax}} \; \big[ p(I \mid S') \; p(S' \mid S, Z) \big],
\end{equation}
where $p(S' \mid S, Z)$ models the refinement process conditioned on both the impaired input $S$ and the latent intention state $Z$. In essence, $Z$ acts as the LLM’s hidden conceptual representation of what the speaker \emph{means}, serving as an internal bridge between acoustic evidence and communicative intent that guides the generation of $S^{\ast}$.


\section{Methodology}

\subsection{Framework Overview}

\begin{algorithm}[ht]
\caption{Speech Refinement Pipeline}
\label{alg:speech_agent}
\begin{algorithmic}[1]
\Require Impaired speech waveform $S \in \mathbb{R}^{l}$, Speaking Style $Style \in \mathbb{R}^{d_s}$
\Ensure Refined text $T'$ and refined speech $S'$
\Statex

\State \textbf{(1) Preprocessing:} Segment $S$ into overlapping frames and compute the log-Mel spectrogram:
\[
\mathrm{MelSpec} \in \mathbb{R}^{F \times \mathcal{T}}.
\]

\State \textbf{(2) Speech Impairment Recognition:}
\State $H_C \gets \mathrm{Encoder}_\theta(\mathrm{MelSpec})$ \Comment{frame-level representation}
\State $h \gets \mathrm{Pool}(H_C)$ 
\State $z \gets W h + b$; \quad $p(y \mid S) = \mathrm{softmax}(z)$
\State $C \gets \arg\max_{y \in \mathcal{C}} p(y \mid S)$ \Comment{predicted impairment class}

\State \textbf{(3) Speech Recognition:}
\State $H_T \gets \mathrm{Encoder}_\phi(\mathrm{MelSpec})$
\State $p(t_i \mid t_{<i}, H_T) = \mathrm{Decoder}(t_{<i}, H_T), \quad t_i \in \mathcal{V}$

\State Impaired text $T = (t_1, \ldots, t_N)$.
\State \textbf{(4) Speech Refinement:}
\State $T' = \mathrm{SpeechRefine}(T, C)$ \Comment{Refine speech with LLM}
\State $P = \mathrm{G2P}(T')$, $X = \mathrm{Embed}(P)$ \Comment{Obtain phoneme embedding}
\State $S' = \mathrm{TTS}(X, Style)$ \Comment{Generate refined speech}

\State \Return $(T', S')$
\end{algorithmic}
\end{algorithm}

\begin{figure*}[htbp]
    \centering
    \includegraphics[width=\linewidth]{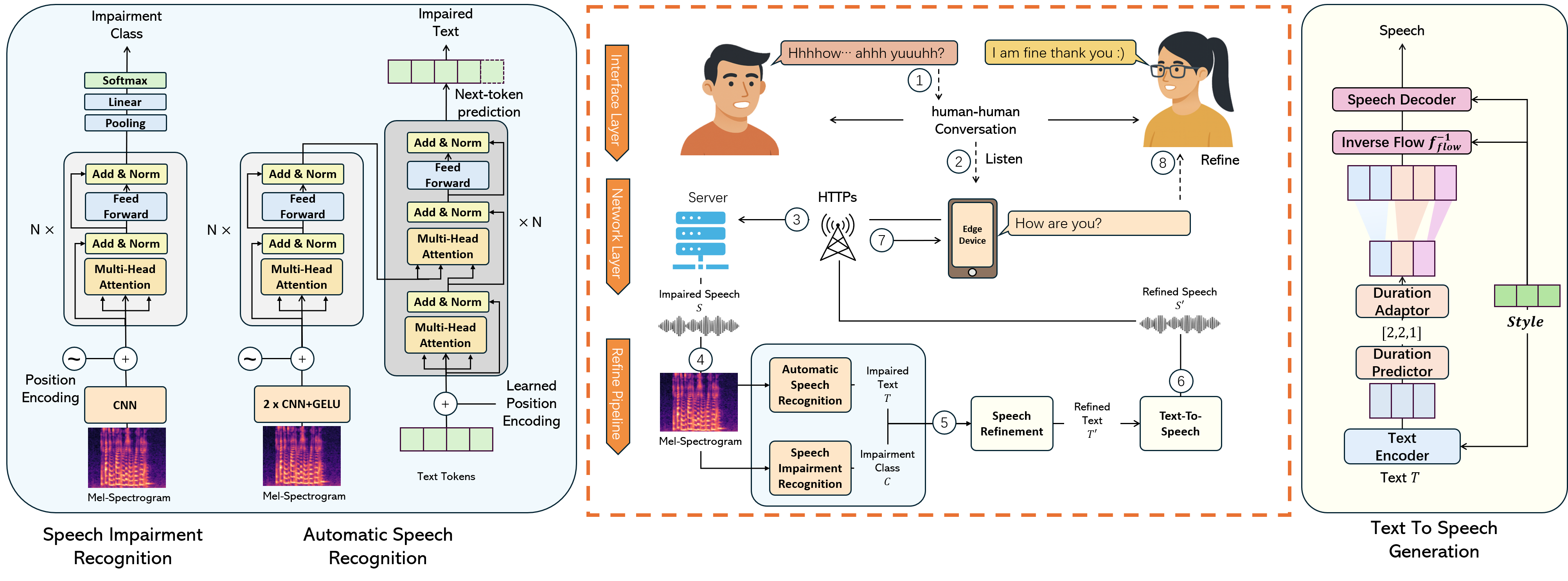}
    \caption{Overview of the system architecture of the proposed \textit{SpeechAgent}. The system integrates speech impairment recognition, LLM-based speech refinement, and text-to-speech generation. It operates in an edge–server setup that refines impaired speech into clearer, intention-preserving output for real-time communication.}
    \label{fig:overview}
\end{figure*}

Our assistive \textbf{SpeechAgent} follows a structured workflow to refine impaired speech. Specifically, given an impaired speech input $S$, the system first identifies the speech impairment class $C$. The signal is then transcribed into text $T$.  At this stage, the raw transcript $T$ may deviate significantly from the true communicative intent $I$. 
We therefore model intent inference as a conditional distribution $P(I \mid T, C)$, where the LLM leverages both the noisy transcript and the impairment class to approximate the underlying intent.  Based on the inferred intent, the system generates a refined transcript $T'$, modeled as $P(T' \mid I, C)$, such that $T' \approx I$. Finally, the refined transcript $T'$ is converted back into speech $S'$, yielding a refined speech that is easier for listeners to understand the intention.  Algorithm \ref{alg:speech_agent} shows the overall workflow.

\subsection{Speech Impairment Recognition}\label{sec:sir_model}
The SIR model gives our \textbf{SpeechAgent} the ability to understand how speech deviates from healthy patterns. Its goal is to detect impairment-related acoustic or linguistic distortions given a speech input. The input to the SIR model is a waveform $S \in \mathbb{R}^l$, represented as a one-dimensional array with $L$ time points. The waveform is segmented into overlapping frames and transformed into a log-Mel spectrogram~$\mathrm{MelSpec} \in \mathbb{R}^{F \times \mathcal{T}}$.
$\mathcal{T}$ is the number of temporal frames and $F$ is the number of frequency bins. The Mel-spectrogram is passed through a seq2seq deep neural encoder $\mathrm{Encoder}_\theta(\cdot)$, parameterized by $\theta$, that processes the input as a sequence of hidden frame-level embeddings $H_C$.:
\begin{equation}
    H_{C} = \mathrm{Encoder}_\theta(\mathrm{MelSpec}), \quad H_C \in \mathbb{R}^{d\times \mathcal{T}},
\end{equation}
To obtain a fixed-length speech-level embedding, we apply a pooling operation over the temporal dimension, $h = \mathrm{pool}(H)$
where $h$ denotes the pooled embedding and $d$ is the hidden dimensionality. In practice, $\mathrm{Pool}(\cdot)$ can be implemented as mean pooling or attention-based pooling over time. This speech-level embedding is then projected into a classification logit:
\begin{equation}
z = W h + b, \quad z \in \mathbb{R}^{|C|},
\end{equation}
where $|C|$ is the number of speech impairment classes. The logits $z$ are converted into a probability distribution over impairment categories using the softmax function:
\begin{equation}
p(y \mid S) = \frac{\exp(z_y)}{\sum_{j=1}^{|C|} \exp(z_j)}, \quad y \in C,
\end{equation}

The predicted class is then given by $\hat{y} = \arg\max_{y \in C} p(y \mid \mathrm{S})$, corresponding to the most likely class of speech impairment.

\subsection{Speech Recognition}
The Automatic Speech Recognition (ASR) model gives our \textbf{SpeechAgent} the ability to ``listen'' and transcribe speech into text. The input to the ASR model is log-Mel spectrogram~$\mathrm{MelSpec}$. 
The Mel-spectrogram is then passed through an seq2seq encoder network $\mathrm{Encoder}_\phi(\cdot)$, parameterized by $\phi$, which maps the input into a sequence of latent embeddings:
\begin{equation}
H_{T} = \mathrm{Encoder}_\phi(\mathrm{MelSpec}), \quad H_T \in \mathbb{R}^{d \times \mathcal{T}},
\end{equation}
where $d$ is the hidden dimensionality. These latent representations capture both short-term phonetic information at the frame level and long-term semantic dependencies across the speech.  On top of this encoder, a decoder network autoregressively generates a sequence of discrete text tokens. At each decoding step $i$, the decoder conditions on the history of previously generated tokens $t_{<i}$ and the latent representation $H$ to produce a probability distribution over the next token:
\begin{equation}
p(t_i \mid t_{<i}, H) = \mathrm{Decoder}(t_{<i}, H), \quad t_i \in \mathcal{V},
\end{equation}

where $\mathcal{V}$ denotes the vocabulary of text tokens. The distribution is typically obtained via a softmax over the decoder output logits. Tokens are generated iteratively until an end-of-sequence symbol is produced. The final transcription is denoted as $T = (t_1, t_2, \ldots, t_N)$, where each $t_i$ corresponds to a token in the recognized text sequence. 

\subsection{Speech Refinement}\label{sec:speech_refine}






Given the impaired text $T$ transcribed by the ASR model and the recognized class of speech impairment $C$, the next step is to enhance its clarity and coherence through a LLM refinement model. This model directly rewrites the transcription into a more fluent and comprehensible form, conditioned on the impairment type $C$. Formally, the refinement can be expressed as :
\begin{equation}
  T' = \mathrm{SpeechRefine}(T, C),
\end{equation}

where $T'$ denotes the refined text. The prompt used for this refinement is presented in Section~\ref{exp:refine_prompt}. Then, the refined text $T'$ is converted to natural speech $S' \in \mathbb{R}^L$. The TTS model provides the agent with the ability to ``speak,'' completing the full reasoning and acting~(ReACT) pipeline. The refined text $T'$ is converted into a phoneme sequence through a grapheme-to-phoneme conversion function $\mathrm{G2P}(\cdot)$~\cite{lou2025generalized}:
\begin{equation}
P = \mathrm{G2P}(T') = [p_1, p_2, \ldots, p_L],
\end{equation}
where $P$ denotes the resulting phoneme sequence and $p_i$ represents the $i$-th phoneme unit.
Each phoneme $p_i$ is then mapped into a continuous embedding vector $x_i \in \mathbb{R}^d$ through a phoneme embedding layer $\mathrm{Embed}(\cdot)$:
\begin{equation}
X = \mathrm{Embed}(P) = [x_1, x_2, \ldots, x_L],
\end{equation}

where $X \in \mathbb{R}^{d \times L}$ denotes the phoneme embedding sequence. We incorporate a speaking style embedding $Style \in \mathbb{R}^{d_s}$, which can be derived from descriptive text prompts, allowing flexible control over the speaking style of generated speech. The TTS model then conditions on both the phoneme embeddings $X$ and the speaking style embedding $Style$ to generate the speech signal:
\begin{equation}
S' = \mathrm{TTS}(X, Style), \quad S' \in \mathbb{R}^L.
\end{equation}

We employ ParaStyleTTS~\cite{lou2025parastyletts}, a neural text-to-speech architecture that jointly models prosody and paralinguistic style for expressive and natural-sounding speech generation. By directly synthesizing waveforms in an end-to-end manner, this TTS model removes the dependency on intermediate spectrogram representations and external vocoders, thereby enhancing both generation quality and computational efficiency.

\section{System Infrastructure}
Our \textbf{SpeechAgent} adopts a client–server architecture \cite{berson1992client}, as illustrated in Figure 2. The mobile edge device acts as the primary user interface, while computation-intensive models are offloaded to a cloud server hosting large-scale models. All communication occurs through secure RESTful APIs over HTTPS, and the server integrates third-party services for monitoring, analytics, and authentication.

\subsection{Edge-Device Functions}
On the client side, the edge device hosts a lightweight application that serves as the primary interface. The application allows users to record speech through the device's microphone and provides simple controls to start or stop interactions with the agent. Once a recording is captured, the audio signal is transmitted to the cloud server over HTTPS~\cite{durumeric2013analysis}. The edge device then receives the processed response, which is synthesized into speech, stored locally, and played back to the user in real time. This design ensures that the mobile client remains efficient and responsive while delegating heavy computation to the cloud server.

\subsection{Server Functions}
The server executes all computation-intensive models through a modular pipeline of models and services. Incoming speech recordings from the mobile app are first processed by a speech-impairment recognition model, which identifies the type of impairment. Next, an ASR system \cite{radford2023robust} transcribes the impaired speech into text. The transcribed text is then refined by LLM \cite{achiam2023gpt, team2024gemini, team2025gemma, yang2025qwen3} that interprets, clarifies, and reformulates the text while preserving semantic fidelity. The refined text is subsequently passed to a TTS model \cite{lou2025generalized}, which converts it into natural, fluent speech. The final audio output is transmitted back to the edge device.
This server-side pipeline, exposed through RESTful APIs, allows seamless integration with the client application and external third-party APIs.


\section{Benchmarking and Evaluation}
Most prior speech impairment-related research has focused on evaluating narrow, well-defined subtasks. For example, studies on speech recognition for impaired speakers typically report word error rate (WER) as the main metric to measure transcription quality~\cite{mirella2024improving,singh2024comprehensive}. Similarly, research on speech impairment detection or classification often relies on standard accuracy or F1 scores to assess model performance~\cite{sheikh2021stutternet}. While these task-specific evaluations provide valuable insights, they do not capture the broader goal of refining impaired speech into a clearer and more communicative form. To systematically benchmark our \textbf{SpeechAgent}, we design an evaluation framework that integrates both speech and text datasets, rigorous preprocessing, and multidimensional performance analysis. Our goal is to capture not only the technical accuracy of impairment recognition and text refinement, but also the practical usability of the system in real-world applications. 


\subsection{Dataset \& Reprocessing}
We employ two categories of datasets to evaluate the performance of the \textbf{SpeechAgent}. First is the impaired speech dataset. To capture diverse speech impairments, we combine multiple sources. For dysarthria, we select 21 hours of English speech from the TORGO dataset~\cite{rudzicz2012torgo}, which contains eight speakers with dysarthria and seven healthy speakers. For stuttering, we select the UCLASS dataset~\cite{howell2009university}, which contains recordings of speakers with stutter. For aphasia, we select speech samples from AphasiaBank~\cite{macwhinney2011aphasiabank}. Additionally, we adopt a pure text dataset, on top of the impaired speech dataset, we select the SNIPS dataset~\cite{coucke2018snips}, a task-oriented text dataset related to common daily-life commands such as weather, transport, reminders, and music. These texts are representative of typical user intents in conversational AI applications and thus provide a practical foundation for evaluating the robustness of the proposed \textbf{SpeechAgent} in refining the impaired speech.

Although numerous studies investigate the automatic recognition of dysarthria~\cite{aboeitta2025bridging}, stuttering~\cite{sheikh2021stutternet}, and aphasia~\cite{azevedo2024artificial}, most approaches treat the task as a binary classification problem, detecting only a single impairment type at a time. Our goal is to equip the \textbf{SpeechAgent} with the ability to recognize multiple impairments simultaneously, which naturally formulates the task as multi-class classification. Due to there is no suitable open-source dataset or model for this setting, we construct our own dataset and train a deep learning model that takes spectrograms as input to classify different types of speech impairment. Specifically, we randomly select \textbf{500} speech samples from each type of speech impairment (dysarthria, stutter, aphasia, and health), and form a dataset with a total of \textbf{2,000} speech samples. The selected speech samples range from 2 to 15 seconds in duration and are resampled to 16 kHz. Each sample is converted into a Mel-spectrogram using a hop size of 256 with a window size of 1,024. The resulting Mel-Spectrogram has approximately \textbf{63} frames per second, where each frame corresponds to approximately \textbf{16ms} of speech signal.  

\begin{table*}[h]
\centering
\caption{Classification performance of different models across impairment classes. Each column group reports Accuracy (Acc), F1-score (F1), and ROC-AUC (AUC). The best performing model is highlighted as \textbf{bold}. All results are averaged over five runs with different seeds, reported as mean $\pm$ standard deviation. }
\label{tab:classification}
\resizebox{\textwidth}{!}
{\begin{tabular}{l|ccc|ccc|ccc|ccc|ccc}
\toprule
 & \multicolumn{3}{c|}{\textbf{Dysarthria}} & \multicolumn{3}{c|}{\textbf{Stuttering}} & \multicolumn{3}{c|}{\textbf{Aphasia}} & \multicolumn{3}{c}{\textbf{Healthy}}  & \multicolumn{3}{c}{\textbf{Overall}}\\
\cmidrule(lr){2-4} \cmidrule(lr){5-7} \cmidrule(lr){8-10} \cmidrule(lr){11-13} \cmidrule(lr){14-16}
\textbf{Model} & Acc & F1 & AUC & Acc & F1 & AUC & Acc & F1 & AUC & Acc & F1 & AUC & Acc & F1 & AUC  \\
\midrule

CLAP & .0$\pm$.0 & .00$\pm$.00 & .19$\pm$.01 & 4.4$\pm$1.4 & .08$\pm$.02 & .44$\pm$.02 & .0$\pm$.0 & .00$\pm$.00 & .16$\pm$.02 & 97.2$\pm$1.2 & .41$\pm$.00 & .36$\pm$.01 & 25.4$\pm$.6 & .12$\pm$.01 & .29$\pm$.02 \\
ParaCLAP & 1.2$\pm$.4 & .02$\pm$.01 & .33$\pm$.02 & 42.0$\pm$5.2 & .20$\pm$.02 & .15$\pm$.02 & 12.0$\pm$1.9 & .17$\pm$.02 & .61$\pm$.02 & 0.4$\pm$0.5 & .01$\pm$.01 & .67$\pm$.04 & 13.9$\pm$2.0 & .10$\pm$.02 & .44$\pm$.02 \\
ASR + LLM & 48.6$\pm$5.3 & .41$\pm$.03 & .59$\pm$.02 & 27.8$\pm$5.4 & .34$\pm$.05 & .58$\pm$.02 & 47.6$\pm$2.3 & .39$\pm$.01 & .58$\pm$.01 & 19.8$\pm$4.3 & .25$\pm$.05 & .54$\pm$.02 & 35.9$\pm$4.3 & .35$\pm$.04 & .57$\pm$.02 \\
CNN + LSTM & 91.2$\pm$3.4 & .92$\pm$.04 & .98$\pm$.01 & 99.8$\pm$.45 & .81$\pm$.02 & .99$\pm$.01 & 67.2$\pm$6.4 & .80$\pm$.04 & .98$\pm$.03 & 86.0$\pm$9.6 & .92$\pm$.06 & \textbf{.99$\pm$.01} & 86.1$\pm$3.6 & .86$\pm$.04 & .98$\pm$.01 \\
Transformer  & \textbf{95.0$\pm$2.0} & \textbf{.95$\pm$.02} & \textbf{.99$\pm$.01} & \textbf{99.8$\pm$.5} & \textbf{.88$\pm$.02} & \textbf{1.00$\pm$.00} & \textbf{82.0$\pm$4.5} & \textbf{.90$\pm$.03} & \textbf{1.00$\pm$.00} & \textbf{88.8$\pm$2.3} & \textbf{.94$\pm$.01} & \textbf{.99$\pm$.01} & \textbf{91.4$\pm$1.7} & \textbf{.92$\pm$.02} & \textbf{.99$\pm$.01}  \\
\bottomrule
\end{tabular}
}
\end{table*}

\subsection{Evaluation}
We evaluate our system across three dimensions to comprehensively assess both its technical performance and practical utility. Specifically, \textbf{recognition accuracy}, which we evaluate whether the \textbf{SpeechAgent} can correctly identify different types of speech impairments;  \textbf{refinement performance}, which we evaluate whether the refined speech by LLM is clear and easy to understand; and \textbf{latency}, which we evaluate whether the communication between the user and the \textbf{SpeechAgent} is sufficiently fast and seamless to support everyday interaction. 

For the classification task, we randomly select 10\% of the speech samples from each impairment class as the test set and use the remaining 90\% for training. A deep learning model is then trained on this dataset, and we report performance in terms of class-level \textbf{accuracy}, \textbf{F1-score}, and \textbf{AUC}. 
These metrics are standard metrics for evaluating classification models, and are widely applied in domains such as facial recognition~\cite{taigman2014deepface} and image classification~\cite{deng2009imagenet}.

For the refinement evaluation, we conduct two types of experiments. The first is a text-based evaluation. We use the text from the SNIPS~\cite{coucke2018snips} dataset as the ground-truth intention and simulate impaired text for each type of speech impairment, using the original text as the reference. The impaired text is then passed through the LLM refinement algorithm. After refinement, we compute semantic similarity between the refined text and the original intention using \textbf{BERT} score~\cite{zhang2019bertscore}, \textbf{cosine score}, and \textbf{BLEU} score~\cite{papineni2002bleu}.  These metrics are widely used to evaluate the quality of language model outputs, including tasks such as machine translation~\cite{vaswani2017attention,papineni2002bleu} and text generation~\cite{zhang2019bertscore}. They measure the similarity between pairs of text; in our context, a higher similarity between the refined text and the original intention indicates that the refined text is clearer to be understood and closer to the true intention, and thus the refinement is more effective.

The second experiment is a speech-based evaluation. We feed the impaired speech samples into an ASR model~\cite{radford2023robust} to obtain transcriptions, which are then processed by the LLM refinement cycle. 
The refined text is subsequently converted back into speech. Since no ground-truth intention is available for the impaired speech samples, we instead conduct a human evaluation. Several human participants are presented with a shuffled list of impaired and refined speech and asked to rate the clarity of the speech and CMOS (Comparison Mean Opinion Score), where we provide the original impaired speech and refined speech to human listeners to judge which one is better. These subjective evaluation methods are widely adopted in the speech generation community for assessing the perceptual quality of TTS models~\cite{wang2017tacotron,shen2018natural,ren2019fastspeech,lou2024stylespeech,lou2025generalized}. In our context, they serve as an effective metric to indicate the overall quality and intelligibility of the refined speech. Additionally, alongside subjective evaluation, we conduct objective evaluation using our trained SIR model. Specifically, we measure the proportion of successfully recovered samples, where a recovery is defined as the SIR model classifying the refined speech as healthy. Lastly, we evaluate the latency of our system to assess its practicality for real-world applications. We build a simulation environment where a edge device communicates with the server to complete the full interaction loop. This process includes transmitting the speech signal, recognizing the impairment type and transcribing the speech, refining the text, generating the refined speech, and transmitting the output back to the user's device. We measure the average end-to-end response time across multiple trials to determine whether the system operates seamlessly and efficiently enough to support everyday user interaction.

\section{Result and Discussion}
In this section, we present our experimental result and discuss them to address the following three research questions (RQs):

\begin{itemize}
    \item \textbf{RO1}: Can the \textbf{SpeechAgent} accurately recognize the class of speech impairment?
    \item \textbf{RQ2}: Can the \textbf{SpeechAgent} effectively refine the impaired speech into a clearer and easier-to-understand form?
    \item  \textbf{RQ3}: Can the \textbf{SpeechAgent} be deployed easily on an edge device so the user can facilitate it for everyday use with low latency? 
\end{itemize}

\begin{table*}[htbp]
\centering
\caption{Text-based refinement evaluation using BERT score, Cosine Similarity, and BLEU. The best score is highlighted in 
\colorbox{green!20}{green}, and the second-best score is highlighted in 
\colorbox{red!20}{red}. All results are averaged over five runs with different seeds, reported as mean $\pm$ standard deviation.}
\label{tab:refinement-text}
\begin{tabular}{l|ccc|ccc|ccc}
\toprule
 & \multicolumn{3}{c|}{\textbf{Dysarthria}} & \multicolumn{3}{c|}{\textbf{Stuttering}} & \multicolumn{3}{c}{\textbf{Aphasia}} \\
\cmidrule(lr){2-4} \cmidrule(lr){5-7} \cmidrule(lr){8-10}
\textbf{Model} & BERT & BLEU  & CosSim & BERT & BLEU  & CosSim & BERT & BLEU  & CosSim \\
\midrule
Impaired & 0.794$\pm$.03 & 0.049$\pm$.04 & 0.045$\pm$.07 & 0.859$\pm$.03 & 0.118$\pm$.07 & 0.819$\pm$.22 & 0.868$\pm$.02 & 0.139$\pm$.10 & 0.412$\pm$.17\\
LanguageTool~\cite{naber2003rule} & 0.858$\pm$.04 & 0.183$\pm$.15 & 0.282$\pm$.19 & 0.862$\pm$.03 & 0.147$\pm$.10 & 0.786$\pm$.22 & 0.866$\pm$.02 & 0.137$\pm$.10 & 0.401$\pm$.17 \\
Gramformer~\cite{gramformer} & 0.801$\pm$.03 & 0.061$\pm$.07 & 0.063$\pm$.10 & 0.872$\pm$.04 & 0.198$\pm$.15 & 0.833$\pm$.21 & 0.870$\pm$.02 & 0.150$\pm$.12 & 0.417$\pm$.17\\
\hline
\multicolumn{10}{l}{\textbf{Refined without impairment class (w/o $C$)}} \\
\hline
GPT-4.1 &  \cellcolor{green!20}0.940$\pm$.05 &  \cellcolor{green!20}0.608$\pm$.32 &  \cellcolor{green!20}0.662$\pm$.30 & 0.961$\pm$.03 &  \cellcolor{green!20}0.771$\pm$.27 &  \cellcolor{red!20}0.861$\pm$.19 &  \cellcolor{green!20}0.915$\pm$.03 &  \cellcolor{red!20}0.278$\pm$.24 &  \cellcolor{red!20}0.458$\pm$.23 \\
Gemini 2.5 & 0.917$\pm$.07 & 0.491$\pm$.39 & 0.533$\pm$.40 & 0.931$\pm$.07 & 0.616$\pm$.41 & 0.659$\pm$.41 & 0.900$\pm$.05 & 0.247$\pm$.25 & 0.387$\pm$.28 \\
Qwen 3 & 0.931$\pm$.06 & \cellcolor{red!20}0.558$\pm$.33 & \cellcolor{red!20}0.612$\pm$.32 & \cellcolor{green!20}0.967$\pm$.04 & \cellcolor{red!20}0.771$\pm$.26 & \cellcolor{green!20}0.861$\pm$.19 & 0.906$\pm$.04 & \cellcolor{green!20}0.281$\pm$.24 & \cellcolor{green!20}0.476$\pm$.23 \\
Gemma3 4B & \cellcolor{red!20}0.935$\pm$.05 & 0.518$\pm$.30 & 0.604$\pm$.30 & \cellcolor{red!20}0.966$\pm$.03 & 0.752$\pm$.27 & 0.856$\pm$.19 & \cellcolor{red!20}0.908$\pm$.03 & 0.254$\pm$.22 & 0.440$\pm$.23 \\
\hline
\multicolumn{10}{l}{\textbf{Refined with impairment class (w $C$)}} \\
\hline
GPT-4.1 & \cellcolor{red!20}0.950$\pm$.04 & \cellcolor{red!20}0.650$\pm$.29 & \cellcolor{red!20}0.709$\pm$.27 & 0.969$\pm$.02 & \cellcolor{red!20}0.815$\pm$.24 & \cellcolor{red!20}0.881$\pm$.18 & \cellcolor{green!20}0.916$\pm$.03 & 0.277$\pm$.23 & 0.456$\pm$.23 \\
Gemini 2.5 & \cellcolor{green!20}0.954$\pm$.04 & \cellcolor{green!20}0.693$\pm$.31 & \cellcolor{green!20}0.745$\pm$.28 & 0.962$\pm$.05 & 0.792$\pm$.30 & 0.833$\pm$.28 & \cellcolor{red!20}0.915$\pm$.03 & \cellcolor{green!20}0.293$\pm$.24 & \cellcolor{red!20}0.465$\pm$.24 \\
Qwen 3 & 0.945$\pm$.05 & 0.611$\pm$.30 & 0.675$\pm$.29 & \cellcolor{green!20}0.974$\pm$.03 & \cellcolor{green!20}0.829$\pm$.22 & \cellcolor{green!20}0.892$\pm$.15 & 0.908$\pm$.04 & \cellcolor{red!20}0.279$\pm$.24 & \cellcolor{green!20}0.472$\pm$.23 \\
Gemma3 4B & 0.941$\pm$.04 & 0.574$\pm$.31 & 0.656$\pm$.29 & \cellcolor{red!20}0.970$\pm$.03 & 0.785$\pm$.25 & 0.873$\pm$.18 & 0.908$\pm$.03 & 0.253$\pm$.23 & 0.437$\pm$.23 \\
\bottomrule
\end{tabular}
\end{table*}

\subsection{Speech Impairment Recognition}
Table~\ref{tab:classification} reports the performance of our speech recognition model. Since there is a lack of prior work directly addressing this multi-modal classification task, we benchmark our approach against two open-source baselines: CLAP~\cite{elizalde2023clap} and ParaCLAP~\cite{jing2024paraclap}, both general-purpose language–audio alignment models. For these baselines, we follow a text–audio similarity approach to make classification. Each speech sample is compared against candidate text prompts of the form \textit{“a speech from [dysarthria, stutter, aphasia, healthy] speaker”}. We then compute the similarity between the speech embedding and each candidate text embedding, and assign the label corresponding to the most similar text.
Additionally, we explore an alternative ASR+LLM baseline. In this approach, the impaired speech is first transcribed using an ASR model, and the resulting transcript is then passed to an LLM, which judges and classifies the impaired speech into one of the speech impairment classes.
In contrast, for the CNN+LSTM and Transformer approaches, we directly train the models to encode Mel-Spectrograms into speech embeddings, which are then optimized using the method described in Section~\ref{sec:sir_model}. A linear classification layer is subsequently applied to predict the speaker condition label from these embeddings. Our trained SIR models will be released upon paper acceptance to encourage reproduction.


Our trained SIR models (CNN+LSTM and Transformer) consistently outperform the baseline methods across all impairment classes. The Transformer model, in particular, achieves nearly perfect recognition with average recognition accuracy over 90\% and ROC-AUC values exceeding 0.99. This superior performance can be attributed to its ability to directly learn acoustic–temporal representations from Mel-Spectrograms, allowing it to capture subtle articulatory and prosodic cues that characterize different impairments. The end-to-end optimization ensures that the learned embeddings are well aligned with the classification objective, resulting in robust and generalizable recognition. These findings indicate that specialized SIR models are essential to achieve reliable classification, as they are better suited to handle the nuanced variability in pathological speech compared to general-purpose alignment models.

In contrast, the baseline approaches exhibit clear limitations. CLAP~\cite{elizalde2023clap} and ParaCLAP~\cite{jing2024paraclap}, while effective in audio-text alignment tasks, struggle in SIR because they were not trained to discriminate pathological speech conditions; their embeddings focus on broad semantic alignment rather than fine-grained acoustic features. Similarly, the ASR+LLM baseline shows moderate improvements but is constrained by the error compounding of two components: ASR mis-transcriptions of impaired speech and the LLM's reliance on textual cues alone, which discard essential acoustic markers of impairment. Together, these results provide strong evidence that dedicated acoustic models are necessary for accurate SIR. Our experimental results demonstrate that the proposed \textbf{SpeechAgent} can accurately recognize speech impairments, which answers \textbf{RQ1}.

\subsection{Refinement from Impaired Speech}
Table~\ref{tab:refinement-text} presents the text-based evaluation for refining impaired text. In this experiment, we mainly evaluate the ability of LLM to reconstruct the intention from the impaired text. Specifically, given an intention text from the Snips dataset~\cite{coucke2018snips}, we first use LLM to synthesize impaired text~$T$ for each type of speech impairment given the intention~$I$ as input (we use GPT-4.1). Then we use different LLMs from various sources to refine the impaired text. Then we compare the semantic similarity between the refined text$T' = \mathrm{LLMRefine(T)}$, with the intention $I$. A higher similarity score indicates better reconstruction of the intention. 
As a baseline, we include grammar correction systems, which represent a lightweight form of text refinement often used in AAC applications to improve intelligibility. We implement both a rule-based grammar correction system (LanguageTool~\cite{naber2003rule}) and a neural grammar correction model (Gramformer, based on~\cite{gramformer}). These serve as strong baselines for comparison with our speech-refinement pipeline.

Results in Table~\ref{tab:refinement-text} demonstrate that LLMs can effectively refine impaired text to better align with the intention. Across all impairment classes, the refined outputs achieve substantial improvements over the impaired baseline in terms of semantic similarity, as reflected by higher BERT score, BLEU, Cosine Similarity, and IMS values. This confirms that LLM-based refinement is capable of recovering much of the speaker's original intention. Among the models evaluated, Gemini 2.5 consistently outperforms other LLMs, achieving the highest refinement performance across all metrics and impairment classes. This indicates stronger robustness in handling diverse impairment patterns. Moreover, incorporating the impairment class ($C$) into the refinement process yields further gains, particularly in challenging cases such as aphasia, where knowledge of the impairment type provides additional context to guide reconstruction. This aligns with our expectation that specifying the impairment condition would help the model focus on relevant linguistic features and thereby improve similarity to the original intention. Hence, this experiment partially demonstrates \textbf{RQ2}, where the LLM can effectively refine the impaired \textbf{text} to a clearer form that is easier to be understood.

\begin{table*}[htbp]
\centering
\caption{Speech-based refinement evaluation. Human participants rate clarity (Likert scale 1–5) and comparison mean opinion score (C-MOS) between impaired and refined speech.}
\label{tab:refinement-speech}
\begin{tabular}{l|ccc|ccc|ccc}
\toprule
 & \multicolumn{3}{c|}{\textbf{Dysarthria}} & \multicolumn{3}{c|}{\textbf{Stuttering}} & \multicolumn{3}{c}{\textbf{Aphasia}} \\
\cmidrule(lr){2-4} \cmidrule(lr){5-7} \cmidrule(lr){8-10}
\textbf{Model} & Clarity & C-MOS & Recover & Clarity & C-MOS & Recover & Clarity & C-MOS & Recover \\
\midrule
Impaired & 1.83  & - & 0.0  & 2.67  & - & 0.0 & 1.83 & - & 0.0  \\
\hline
ASR + TTS  & 4.67 & 1.83 & 60.0  & 4.00  & 0.83 & 10.0 & 3.33 & 1.17 & 20.0\\
AutomaticCorrection~\cite{arjun2020automatic} & 1.17  &-1.00  & 10.0  & 1.33  &-0.50  & 0.0 & 1.67 & -0.67 & 10.0  \\
CleanVoice~\cite{cleanvoice2025stutter} & 1.50  & 0.00 & 0.0  &2.50  & -0.17 & 0.0 & 1.67 & 0.00 & 0.0  \\
LanguageTool~\cite{naber2003rule} & 4.50  & 1.83 & 50.0  & 4.17  &0.67  & 10.0 & 3.67 & 1.17  & 20.0\\
Gramformer~\cite{gramformer} & 4.83 & 1.83  & 50.0  & 4.17 & 1.17 & 0.0 & 3.83 & 1.33  & 40.0\\
\hline
SpeechAgent & 5.00  & 1.83 & 60.0 & 4.50 & 1.33  & 10.0 & 4.83 & 2.00 &  40.0  \\
\bottomrule
\end{tabular}
\end{table*}

Table~\ref{tab:refinement-speech} presents the speech-based evaluation for refinement of impaired speech. In this experiment, we evaluate the ability of the \textbf{SpeechAgent} to refine the impaired speech in an end-to-end manner. Given an impaired speech $S$, we pass it through the entire \textbf{SpeechAgent} and let the \textbf{SpeechAgent} refine and produce the refined speech~$S'$. 
Purely acoustic-based baselines~\cite{cleanvoice2025stutter,arjun2020automatic}, which focus on signal-level modifications such as repetition removal or silence trimming, achieve only marginal or even degraded performance compared to the original impaired speech. This decline arises because these approaches overlook semantic consistency. The impaired speech signal is already challenging to interpret, and further acoustic alterations often distort meaning and naturalness. Text-based baselines employing grammar correction~\cite{gramformer,naber2003rule} perform more competitively, particularly for dysarthric speech, where linguistic structure remains intact but articulation is weak. In such cases, accurate transcription alone (ASR + TTS) already yields speech of comparable clarity to our full agent, indicating that dysarthric speech primarily suffer from articulatory, not semantic, impairment. In contrast, grammar correction performs poorly on stuttering and aphasia, as these impairments introduce disruptions or omissions at the semantic level that rule-based correction cannot resolve. LLMs, however, can reconstruct the underlying intent beyond syntactic adjustment, producing refined text that is both fluent and semantically easy to be understood. These performance gains are particularly evident in the stuttering and aphasia scenarios, where the LLM effectively removes repetitions, restores disrupted sentence structures, and reconstructs missing or fragmented content to improve overall clarity.  Taken together, these findings confirm that the \textbf{SpeechAgent} is effective at enhancing the clarity of impaired speech, thereby directly addressing \textbf{RQ2}.

\subsection{Efficiency Analysis}
\begin{figure}[tb]
    \centering
\includegraphics[width=0.45\linewidth]{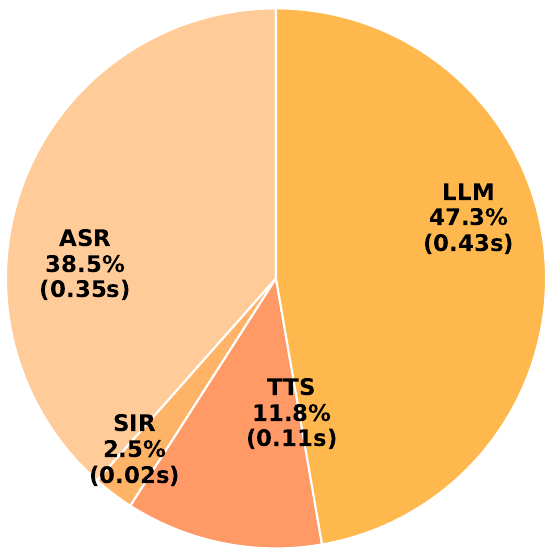}
    \caption{Model-wise inference time for refining speech samples with average duration of 11.49s}
    \label{fig:inference_time}
\end{figure}

\begin{figure}[htb]
    \centering
    \includegraphics[width=0.9\linewidth]{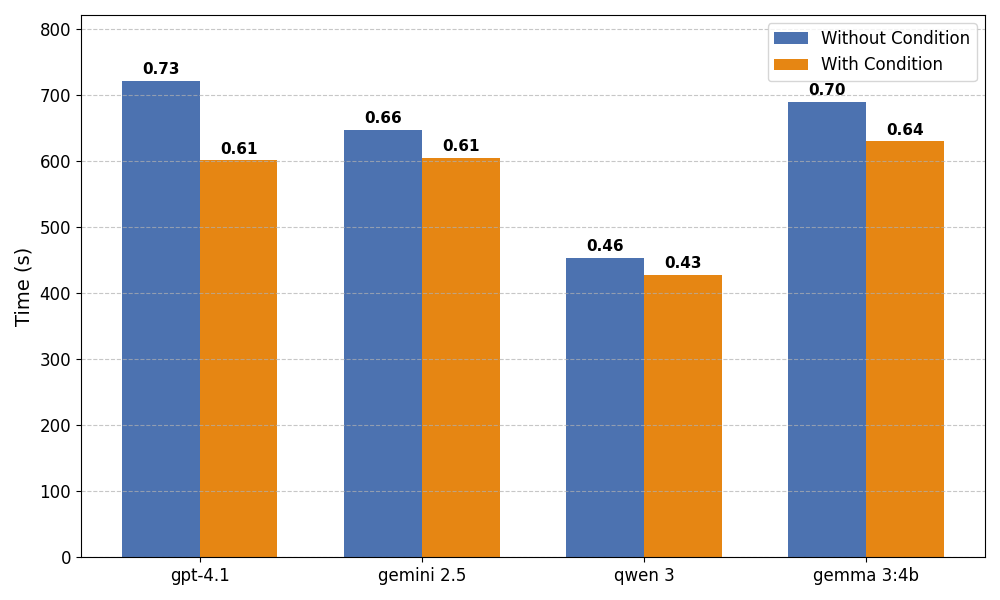}
    \caption{LLM refine time comparison}
    \label{fig:llm_inference_time}
\end{figure}

Lastly, we evaluate the efficiency and latency of the proposed \textbf{SpeechAgent}. 
We report both \textit{refinement latency} (comparison between different LLMs to complete a refinement) and \textit{model-wise latency} (ASR, SIR, Speech refinement, and TTS).  Our experimental result is presented in Figure~\ref{fig:llm_inference_time} \& \ref{fig:inference_time}. The ASR and LLM models account for most of the inference time, with the speech recognition model taking \textbf{38.5\%} and the LLM refinement model taking \textbf{47.3\%}. Nevertheless, the overall speech refinement pipeline achieves near real-time performance across all components. On average, refining a sentence of approximately 11.5 seconds in duration requires only \textbf{0.91 seconds} of processing time, yielding a real-time factor (RTF) of \textbf{0.08}, far below the real-time threshold (RTF = 1). It keeps the latency well within the acceptable range for natural human communication~\cite{jacoby2024human}. This indicates that the proposed approach is practically viable for real-world use, enabling fluid and responsive interaction. Consequently, the efficiency evaluation provides a clear answer to \textbf{RQ3}, demonstrating that the \textbf{SpeechAgent} can reliably meet the latency requirements of everyday conversations.

\section{Conclusion}

In this paper, we presented a mobile \textbf{SpeechAgent} that integrates the reasoning capabilities of LLMs with advanced speech processing to enable real-time assistive communication for individuals with speech impairments. Our system demonstrates that low-latency and practically deployable assistive \textbf{SpeechAgent} can be realized through a carefully designed client–server architecture. By combining impairment recognition, adaptive refinement, and seamless mobile integration, the agent delivers personalized, intelligible, and context-aware communication support. Evaluation on real-world data confirms both technical robustness and positive user experience, suggesting that the proposed approach is feasible and effective in everyday use. Looking ahead, several directions remain open for exploration. While the current system performs reliably, occasional hallucinations from the language model can lead to subtle deviations from the speaker's intention. Future work could focus on enhancing semantic consistency and intent preservation. Moreover, the present agent primarily serves as a passive assistant that refines impaired speech into clearer expressions. Extending its capability toward a more interactive, training-oriented role could transform it into an adaptive partner that helps users practice, monitor, and improve their speech over time. Future research may also explore multilingual and personalized extensions, enabling the agent to support diverse linguistic, emotional, and contextual variations while maintaining natural and trustworthy communication.

\bibliographystyle{ACM-Reference-Format}
\bibliography{custom}

\appendix
\section{Questionnaire}


This section outlines the perceptual evaluation protocol for assessing refined speech quality. Two complementary rating schemes were used to capture absolute and relative listener judgments. The Clarity (1–5) score measures overall intelligibility, considering pronunciation, fluency, and ease of understanding. The Comparison Mean Opinion Score (C-MOS, –3 to +3) assesses the perceived improvement from the original to the refined speech, reflecting how effectively the refinement enhances clarity and fluency.

\noindent \textbf{Clarity (1–5)}
This score evaluates overall speech clarity, considering both the acoustic quality (pronunciation, fluency, ease of listening) and the semantic quality (sentence flow, coherence, and whether the speaker's intent is understandable). 
\begin{itemize}
\item \textbf{1:} Very unclear—speech is poorly articulated or highly disfluent; meaning is obscured by ambiguities or contradictions.
\item \textbf{2:} Somewhat unclear—noticeable issues in pronunciation or flow; listener must exert significant effort to grasp meaning.
\item \textbf{3:} Moderately clear—speech and intent are understandable, though pronunciation or structure could be refined for smoother comprehension.
\item \textbf{4:} Clear—speech is well articulated and sentences flow logically; intent is easy to follow with minimal effort.
\item \textbf{5:} Very clear—speech is fluent, precise, and natural; intent is conveyed effortlessly and effectively.
\end{itemize}

\noindent \textbf{Comparison Mean Opinion Score (C-MOS, –3 to +3)}
In this task, you will be presented with two speech samples. Please listen carefully to both and rate how the refined speech compares to the original.  

\begin{itemize}
    \item \textbf{–3:} Sample B is much worse than Sample A (severely reduces clarity or quality).  
    \item \textbf{–2:} Sample B is moderately worse than Sample A.  
    \item \textbf{–1:} Sample B is slightly worse than Sample A.  
    \item \textbf{0:} No noticeable difference between the two samples.  
    \item \textbf{+1:} Sample B is slightly better than Sample A.  
    \item \textbf{+2:} Sample B is moderately better than Sample A.  
    \item \textbf{+3:} Sample B is much better than Sample A (significant improvement in clarity, naturalness, or communicative effectiveness).  
\end{itemize}

\noindent\textbf{Instructions to Participants:}  
For each question, you will hear two speech samples (A and B).  
Please compare them and select a score from –3 to +3, according to the scale above.

\section{Case Study}
Case study for three types of speech impairments. The corresponding samples can be found in the provided anonymous Git repo.

\begin{tcolorbox}[colback=orange!2, colframe=orange!40, title=\textbf{Dysarthria}]
\textcolor{examplecolor}{\textbf{E1:} Brr... aaigh... s-sun... shhhiiin... (pause) shhh...imm-murrs... onn... thuhh... ohh-shhhn..}\\
\textcolor{refinedcolor}{\textbf{R1:} Bright sunshine shimmers on the ocean.}\\[2pt]

\textcolor{examplecolor}{\textbf{E2:} Yuu'd... b–buhh... bett...uhh... aa–sss... ta–kinn... uh... cohhl'd... shhh–ow–uhh...}\\
\textcolor{refinedcolor}{\textbf{R2:} You'd be better off taking a cold shower.}\\[2pt]

\textcolor{examplecolor}{\textbf{E3:} Ghh...iiiv...innn... thooss... wuhh... obb–zzerrv... hhhimm... uh... prr–nnaunnnss'd... fee–linn... uhv... thuhh... uhhht...mohhst... rr–ehhss...sp–eckt...s.}\\
\textcolor{refinedcolor}{\textbf{R3:} Giving those who observe him a pronounced feeling of the utmost respect.}\\
\end{tcolorbox}


\begin{tcolorbox}[colback=orange!2, colframe=orange!40, title=\textbf{Stutter}]
\textcolor{examplecolor}{\textbf{E1:} Lotss, it's got five diff–different stages... it's... it's time attack, it's survival, it's practice mode... there's mode, ... there's and there's option mode, and there is}\\
\textcolor{refinedcolor}{\textbf{R1:} I think it got five different stages, there is time attack, there is survival, there is practice mode, there is versus mode, there is option mode and there is...}\\[2pt]

\textcolor{examplecolor}{\textbf{E2:} A–at this Friday... I–I'm going with my mum, my two b–brothers, my s–sister... and one of my c-cousin... and we're all g-going there... we're all going there.}\\
\textcolor{refinedcolor}{\textbf{R2:} This Friday, I'm going with my mum, my two brothers, my sister, and one of my sisters. We're all going there together.}\\[2pt]

\textcolor{examplecolor}{\textbf{E3:} ...S–(siris?)... and th–this is... this is g–generally about... s–sss... it's about a scientist who has... who has...}\\
\textcolor{refinedcolor}{\textbf{R3:} ...Series, and it is generally about a scientist who has...}\\
\end{tcolorbox}


\begin{tcolorbox}[colback=orange!2, colframe=orange!40, title=\textbf{Aphasia}]
\textcolor{examplecolor}{\textbf{E1:} When I r–realized I'd had the stroke... I w–wanted to go to Emerge... 'cause I knew there was... some... [//] a w–window... you could get for the... [/] uh... uh... the b–blood.}\\
\textcolor{refinedcolor}{\textbf{R1:} When I realized I had the stroke, I wanted to go to the emergency room because I knew there was a window for getting treatment for the blood.}\\[2pt]

\textcolor{examplecolor}{\textbf{E2:} Uh–um... I... I w–was... im... important to... m–m... m–p... p... me to... (pause)}\\
\textcolor{refinedcolor}{\textbf{R2:} It was important to me to...}\\[2pt]

\textcolor{examplecolor}{\textbf{E3:} And... (be)cause he said... pro–prob... probably the... t–t... uh... t–te... uh... t–t... t... el... s... ti...}\\
\textcolor{refinedcolor}{\textbf{R3:} And because he said probably the test.}\\
\end{tcolorbox}

\section{User Interface}
The user first taps the record button (Figure 5a) to start a session, and the mobile device records speech until the user presses stop (Figure 5b). The recorded speech is then sent to the server for transcription via the ASR model (Figure 5c), after which the LLM refines the text (Figure 5d). The text is converted to speech, transmitted to edge device and assists user for communication (Figure 5e).
\begin{figure}[h]
    \centering
    \begin{subfigure}[b]{0.26\linewidth}
        \centering
        \includegraphics[width=\linewidth]{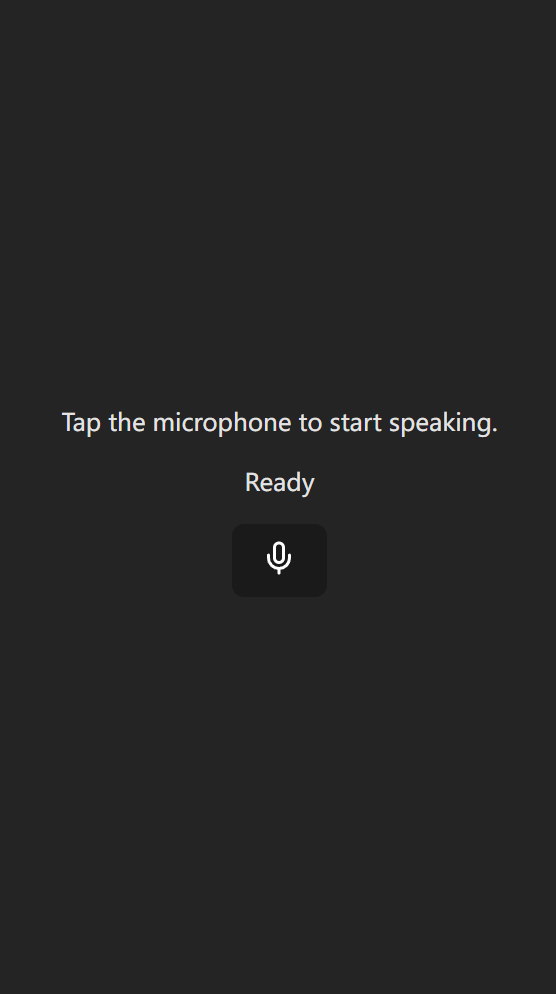}
        \caption{Start Session}
    \end{subfigure}
    \hfill
    \begin{subfigure}[b]{0.26\linewidth}
        \centering
        \includegraphics[width=\linewidth]{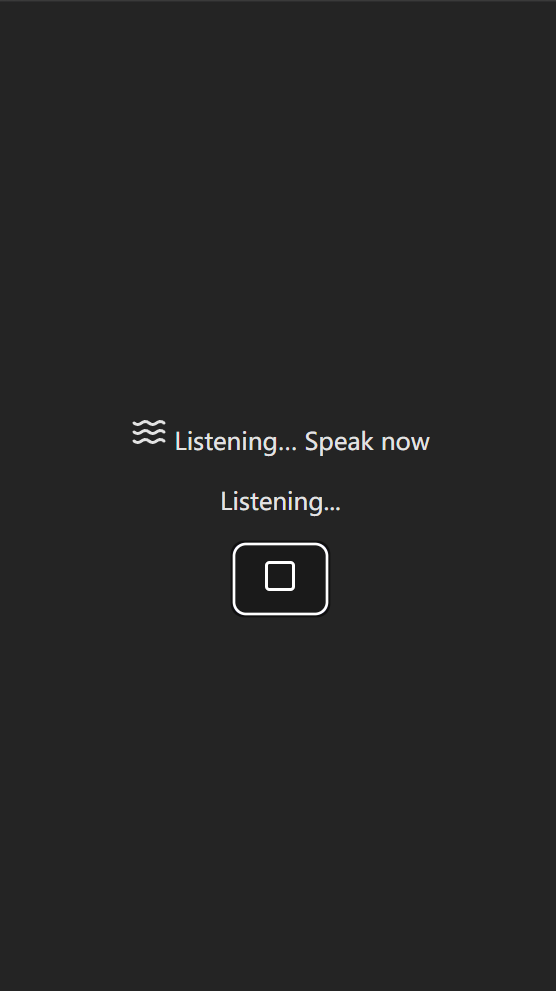}
        \caption{User speak}
    \end{subfigure}
    \hfill
    \begin{subfigure}[b]{0.26\linewidth}
        \centering
        \includegraphics[width=\linewidth]{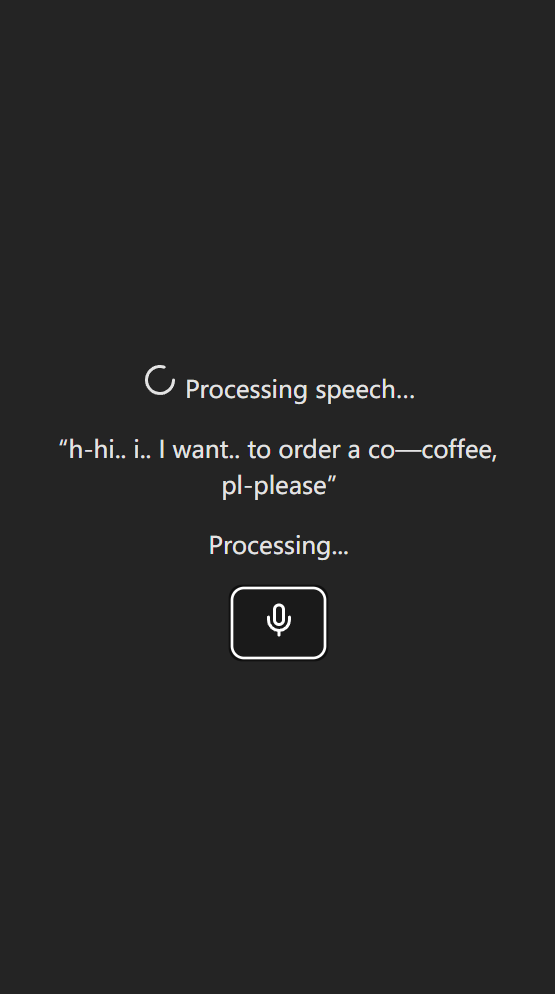}
        \caption{Transcript}
    \end{subfigure}
    \hfill
    \begin{subfigure}[b]{0.26\linewidth}
        \centering
        \includegraphics[width=\linewidth]{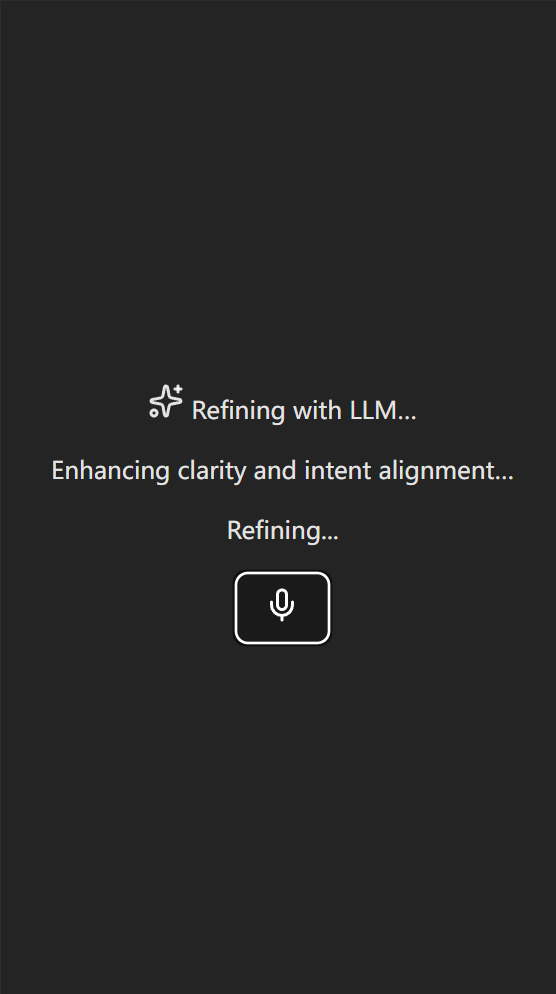}
        \caption{Refining speech}
    \end{subfigure}
    \hfill
    \begin{subfigure}[b]{0.26\linewidth}
        \centering
        \includegraphics[width=\linewidth]{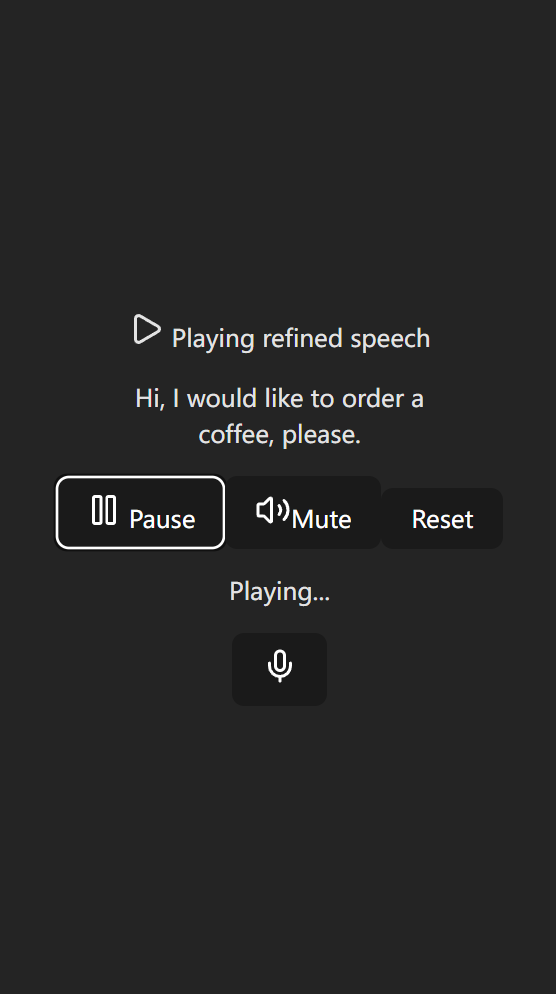}
        \caption{Assisst Communication}
    \end{subfigure}
    \caption{Mobile \textbf{SpeechAgent} UI Demo. }
\end{figure}


\section{Prompt Design}
This prompt defines how the large language model (LLM) performs speech refinement for impaired speech in Section~\ref{sec:speech_refine}. Two variants are used: one without explicit impairment context and another conditioned on the detected impairment type. In both cases, the LLM reconstructs incomplete or unclear sentences into coherent and fluent expressions while preserving the speaker’s original intent. When an impairment class is provided, additional contextual cues (e.g., slurred articulation in dysarthria, repetition in stuttering, or lexical omission in aphasia) guide the refinement process, enabling the LLM to adapt its rewriting strategy to different impairment patterns. 

\begin{tcolorbox}[colback=myred!10, colframe=myred!100, title=\textbf{LLM Refine Prompt}]\label{exp:refine_prompt}
\small
\textbf{Refinement without Impairment Class:}

\begin{verbatim}
You are a helpful assistant that refines 
incomplete or unclear sentences into clear, 
complete expressions. Your task is to rewrite 
the sentence by inserting or clarifying the 
missing content, using the supplement to 
guide your reconstruction. Keep the meaning 
close to the input. 
Input: [impaired text]
Output:
\end{verbatim}

\vspace{6pt}
\textbf{Refinement with Impairment Class:}
\begin{verbatim}
You are a helpful assistant that refines 
incomplete or unclear sentences into clear, 
complete expressions. Your task is to rewrite 
the sentence by inserting or clarifying the 
missing content, using the supplement to 
guide your reconstruction. Keep the meaning 
close to the input. 
Condition: [impairment description]
Input: [impaired text]
Output:
\end{verbatim}

\textbf{Impairment Description:}

\textbf{Dysarthria:} this is a text from a speaker with dysarthria, who may have slurred or slow speech that can be difficult to understand. The text may contain mispronunciations or phonetic errors. \\[4pt]
\textbf{Stutter:} this is a text from a speaker with a stutter, who may have repetitions of sounds, syllables, or words, as well as prolongations and blocks. The text may reflect these speech disruptions. \\[4pt]
\textbf{Aphasia:} This is a text from a speaker with aphasia, who may have difficulty finding the right words or constructing sentences. The text may contain omissions, substitutions, or jumbled words.

\end{tcolorbox}

\end{document}